# MRI Images Analysis Method for Early Stage Alzheimer's Disease Detection

**Achraf Ben Miled [1], Taoufik Yeferny [1] and Amira ben Rabeh [2]**

[1]Faculty of Science, Northern Border University, Saudi Arabia;
[2]SEU Saudi Electronic University, Saudi Arabia;

**Summary**
Alzheimer's disease is a neurogenerative disease that alters memories, cognitive functions leading to death. Early diagnosis of the disease, by detection of the preliminary stage, called Mild Cognitive Impairment (MCI), remains a challenging issue. In this respect, we introduce, in this paper, a powerful classification architecture that implements the pre-trained network AlexNet to automatically extract the most prominent features from Magnetic Resonance Imaging (MRI) images in order to detect the Alzheimer's disease at the MCI stage. The proposed method is evaluated using a big database from OASIS Database Brain. Various sections of the brain: frontal, sagittal and axial were used. The proposed method achieved 96.83% accuracy by using 420 subjects: 210 Normal and 210 MRI.

## 1. Introduction

Chronic Alzheimer's disease is one of the most insidious diseases of modern society. In 2019, 40 million people were affected by this disease [1,2]. In fact, treatment of the latter is extremely expensive for health care systems around the world [3]. Although there is no way to stop the disease when it is too advanced, studies show that its progression can be slowed down or interrupted if it is identified at an early stage [4,5]. Alzheimer's is the most common neurodegenerative disease among the elderly. Early diagnosis is crucial both to treat the disease and to assist in the development of new drugs. Up to now it has not been possible to find a cure [6,7]. The development of Alzheimer's disease has been very closely related to changes in the white matter, and with regard to gray matter. The latter is responsible for the treatment of information, and the function of the white substance. It connects the different regions of the brain by fibers of cerebral connectivity network. A significant loss of fiber causes functional disorders such as memory loss [8,9,10,11].

The diagnosis Alzheimer's disease remains a challenge. Despite scientific progress, it has not yet been possible to determine how functional brain activity deteriorates the structure and vice versa. This is a key element to better understand the development of this type of disease [12,13]. One of the difficulties of Alzheimer's disease is that by the time all the clinical symptoms appear, and a definitive diagnosis is established, too many neurons are dead. This makes the disease essentially irreversible. In this regard, Computer-Assisted Diagnostics (CAD) is an important tool that helps physicians to interpret the multimedia content obtained by testing patients, which makes the application of treatment easier. Among these methods are medical images, which provide high resolution "in vivo" information on the study subjects. It allows the use of information related to the disease shown in the image. However, early diagnosis of Alzheimer's disease is extremely important, as treatments are more effective at early stage of the disease. To this end, "Deep Learning" methods could be of use to detect the signs of diseases since its higher performance compared to that of human. Deep Learning (DL) is a subset of Artificial Intelligence (AI) and Machine Learning (MI) based on multilayer artificial neural networks. It provides high precision in several fields such as object detection, speech recognition, language translation, to cite but few. In this respect, we introduce in this paper a deep learning algorithm that implements AlexNet architecture for detecting Alzheimer's disease at early stage through MRI images analysis.

The rest of the paper is organized as follows. In Section 2, we introduce the related work. Section 3 thoroughly discusses our solution. In Section 4, we report the results of our experimental evaluation. Section 5 concludes and sketches issues for future work.

## 2. Related Work

The literature witnesses several methods that have tried to detect Alzheimer's disease. Saruar Alam et al [15] proposed a method that differentiates between Alzheimer's and simple control. Using dual-tree complex wavelet transforms (DTCWT) and the machine learning model SVM (Support Vector Machine). In fact, they chosed the axial section as the type of MRI. Other researchers have worked on deep learning architectures: Convolutional neural networks were used for the classification of Alzheimer's disease and the differentiation between MCI and Alzheimer's disease [16,17,18,19,20,21,22,23]. Convolutional Neural Networks (CNNs) are a class of deep learning networks that dominate in the field of computer-assisted diagnostics based on models including AlexNet [24], ZF Net [25], VGG [26] and GoogleNet [27].

All transfer learning methods used the ImageNet dataset, which is database of annotated images produced by the ImageNet organization, intended for research work in computer vision. Jain et al, proposed a deep transfer learning method-based Alzheimer's disease diagnosis system [29]. They built a new computer assisted diagnosis. The authors used the VGG-16 network [26]. This network consists of several layers, including 13 convolutional layers and 3 fully connected ones. It must therefore learn the weights of 16 layers. The proposed method takes as input a colored image of size 224×224 then classifies it in one of the 1000 classes. It therefore returns a vector of size 1000, which contains the probabilities of belonging to each of the classes. The architecture of VGG-16 is illustrated by the diagram shown in Figure 1.

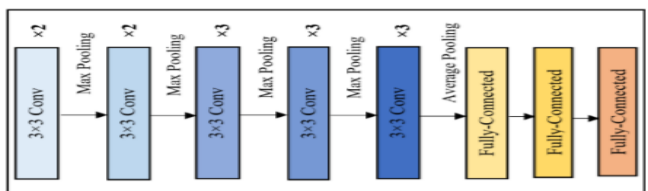

Figure 1. Architecture of the network VGG16

There are four types of layers for a convolutional neural network: the convolution layer, the pooling layer, the ReLU correction layer and the fully-connected layer. The convolutional layer is the key component of convolutional neural networks, which always constitutes at least their first layer. Its purpose is to identify the presence of a set of features in the images received as input. To do this, we perform convolution filtering: the principle is to "drag" a window representing the feature on the image and calculate the convolution product between the feature and each portion of the scanned image. A feature is then seen as a filter.  Pooling layer is the type of layer placed between two convolutional layers: it receives as input several features then applies the pooling operation to each of them. The pooling (or sub-sampling) operation consists in reducing the size of the images, while preserving their important characteristics. The pooling layer makes it possible to reduce the number of parameters and calculations in the network. This improves the efficiency of the network and avoids over-learning. The ReLU layer improves the efficiency of the processing by inserting between the processing layers a new layer which will apply a mathematical function (activation function) on the output signals. The ReLU correction layer therefore replaces all negative values received as inputs with zeros. The fully-connected layer is always the last layer of a neural network, (i.e., convolutional or not). This type of layers receives a vector as input then produces a new output vector. To do this, it applies a linear combination and then optionally an activation function to the values received as input. The fully-connected layer allows to classify the image at the input of the network: it returns a vector of size N, where N is the number of classes in our image classification problem. Each element of the vector indicates the probability for the input image to belong to a class.

Atif Mehmood et al proposed a new system to detect Alzheimer in early phase [30]. The authors proposed to detect Alzheimer disease using the network googlenet. The latter is composed of several layers. They are composed of several convolution modules of size $1 \times 1$, $3 \times 3$ and $5 \times 5$, executed in parallel on the characteristics card resulting from the previous layer.  The architecture of GoogleNet is illustrated by the diagram shown in Figure 2.

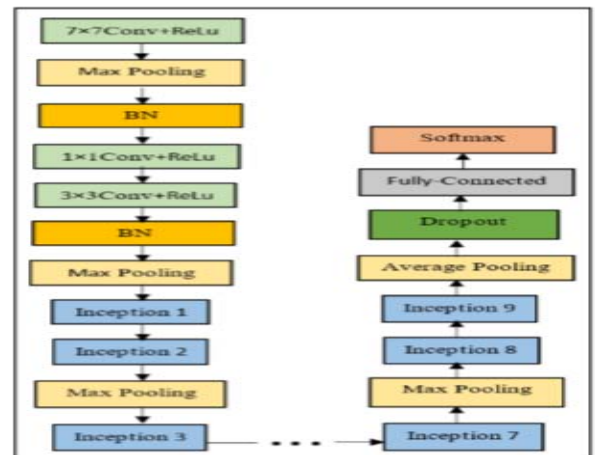

Figure 2. Architecture of the network GoogleNet

Farheen Ramzan et al proposed [31], a new technique to predict AD. The authors used the ResNet18 network that consists of several layers, including 16 convolutional layers and 2 fully-connected. The architecture of ResNet is illustrated by the diagram depicted in Figure 3.

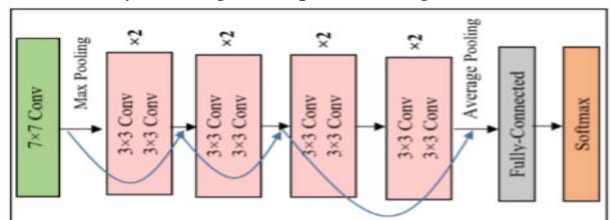

Figure 3. Architecture of the network ResNet18

The existing methods can detect Alzheimer or differentiate between Alzheimer and normal subject. However, they didn't consider differentiating between the MCI and the normal subject. The MCI stage is an intermediate stage between normal state and dementia, which gives the patient a significant risk of progressing towards a stage of dementia in the subsequent years. To overcome this shortness, we propose a new and efficient method that distinguishes between MCI and normal subject.

## 3. Proposed Method

### 3.1. Materials

We have access to more than 420 MRI images from OASIS Database Brain. The learning algorithm performed on 70% of this dataset; the remaining 30% of the dataset was used to test the AI.

### 3.2. Methodology

We used the technology of transfer learning with the performed network Alexnet. It is possible to transfer capability thanks to the increasingly complex hierarchical representations obtained during the learning phase. Indeed, the first layers of neural networks learn to look for very simple characteristics in images (lines, curves, colors, etc.). These latter are largely independent of the problem to be solved, so they are relevant and reusable for a large number of different tasks. When we apply the Transfer Learning, we cut the part of the network specific to a particular task and we keep the general part. The main contributions are:

- The use of the technology of transfer learning to differentiate between MCI and Simple control with the preformed network Alexnet (just 5 Convolutional layers) in order to accelerate the treatment.
- The use of many slices of the brain: frontal, sagittal and axial; a various slice to analyze the brain in many orientations and to detect all component inside.

Figure 4 depicts the model of the proposed method. In fact, to analyze images in a meaningful way, it is important to analyze the hierarchy of objects defined by pixels rather than each pixel separately. Convolutional neural networks analyze images since they use layers of neurons called convolutional layers. These layers of neurons scan the image to detect its pertinent features (for the problem to be solved). The superposition of several convolutional layers makes possible the hierarchical detection of characteristics that are more complex in an image.

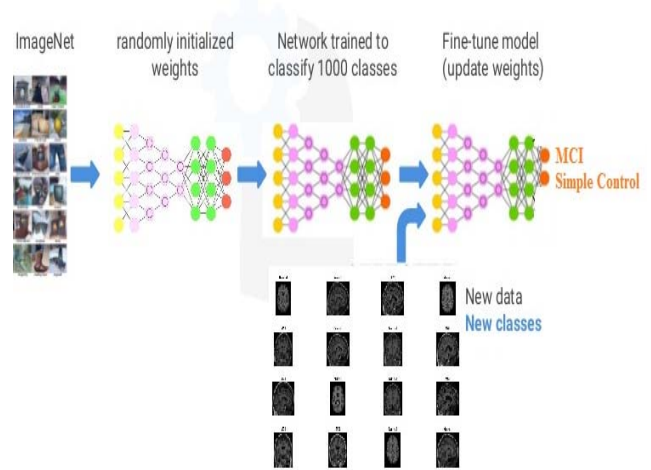

Figure 4. Model of the proposed method

We rely on AlexNet which is the neuron network that won ImageNet in 2012. It was a deep convolutional neural network composed of 5 convolutional layers and three more classical neural layers. The most commonly used cases in image analysis are the classification of images (find the category of the main object of the image), the detection of objects (find where all the objects of the image are located), image and image segmentation (determine for each pixel of the image which category it belongs to). These three used cases are increasingly complex. However, with enough data, current neural networks can achieve similar performance to humans.

This Artificial Intelligence (AI) technique is designed to model high-level abstractions into data to ensure that computers learn to differentiate the brain of a healthy person from that of a sick person by extracting automatically the regions of interest that are affected. In this study, we rely on deep learning techniques to compute predictors on images of brain function and magnetic resonance to prevent Alzheimer's disease.

Most Deep Learning applications use the Transfer Learning method to develop a pre-trained model. The process starts with an existing network, such as AlexNet or GoogLeNet that requires adding new data containing classes previously unknown to the network. Once some adjustments to the network have made, a new task, such as categorizing dogs or cats, rather than 1,000 different objects can be performed. This technique has also the advantage of requiring a much smaller amount of data (thousands of images must be processed, rather than millions), making the computation a matter of minutes rather than hours.

## 4. Results and Discussion

The proposed method achieved a score around 96.83% in the detection of the disease at an early stage. The precision rates obtained by the diagnosis make it possible to advance greatly in the knowledge of the neurodegenerative process involved in the development of the disease, as well as to serve as a starting point for the development of more effective medical treatments to Alzheimer's.

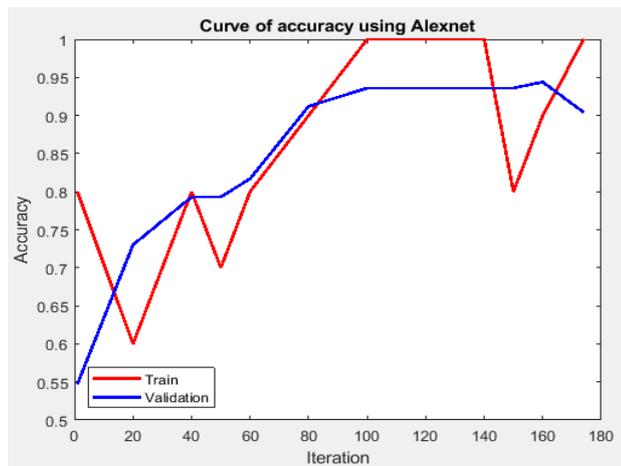
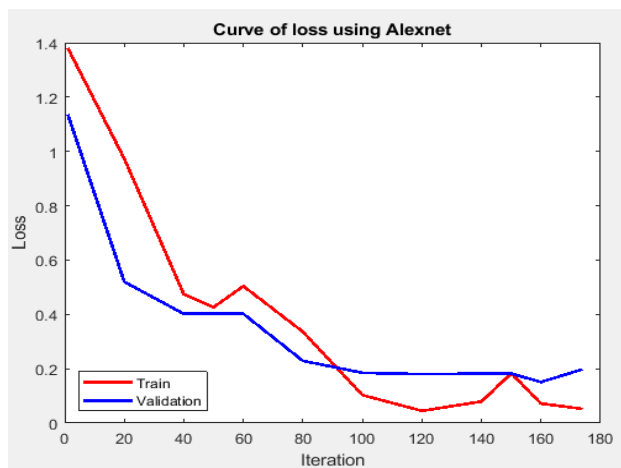

Figure 5. Obtained results using Alexnet

Whenever Alzheimer's disease is diagnosed where all the symptoms have occurred, the volume of the brain is so large then it is too late to intervene, and the disease go up. Detecting the disease at an early stage allows to find a way for slowing down or even stopping the disease. Figure 10 depicts some predictive results detecting the early stage of Mild Cognitive Impairment (MCI) of Normal Control. We evaluated the accuracy of the MRI images classification within our method compared to three pioneering baselines. The achieved accuracies are shown in Figures 5-9. In fact, the four assessed methods AlexNet, VGG16, GoogleNet and ResNet18 achieve an accuracy around 96.83%, 93.95%, 90.4% and 83.33%, respectively.

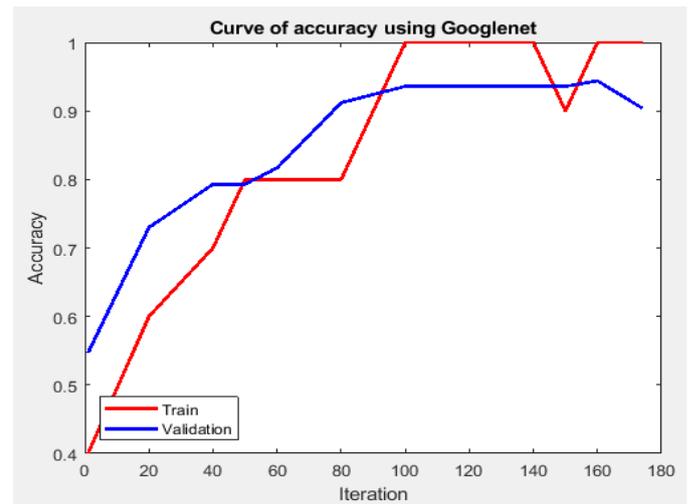
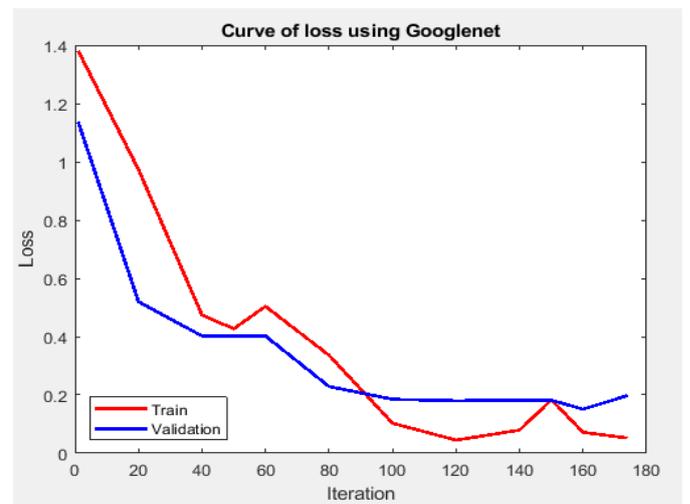

Figure 6. Obtained results using Googlenet

This study included more patients with other cognitive impairments that can be at the origin of transfer learning offers the most useful features. The manual engineering of data is therefore much less when using Deep Learning rather than more traditional machine learning algorithms. This makes it possible to take full advantage of the very large data volumes and computing power that are now available since the emergence of Big Data.

Unlike more traditional algorithms, which offer learning solutions that work for only one problem, the solutions found by deep neural networks can often be applied to other similar tasks (with some adaptations). This is called Transfer Learning. For example, it is possible to use neural networks that have been learned from ImageNet competition images and to re-adapt them so that they can recognize new types of objects. This method saves time since it is not necessary to learn everything from the network, the knowledge base allowing it to distinguish one image from another that has already been learned.

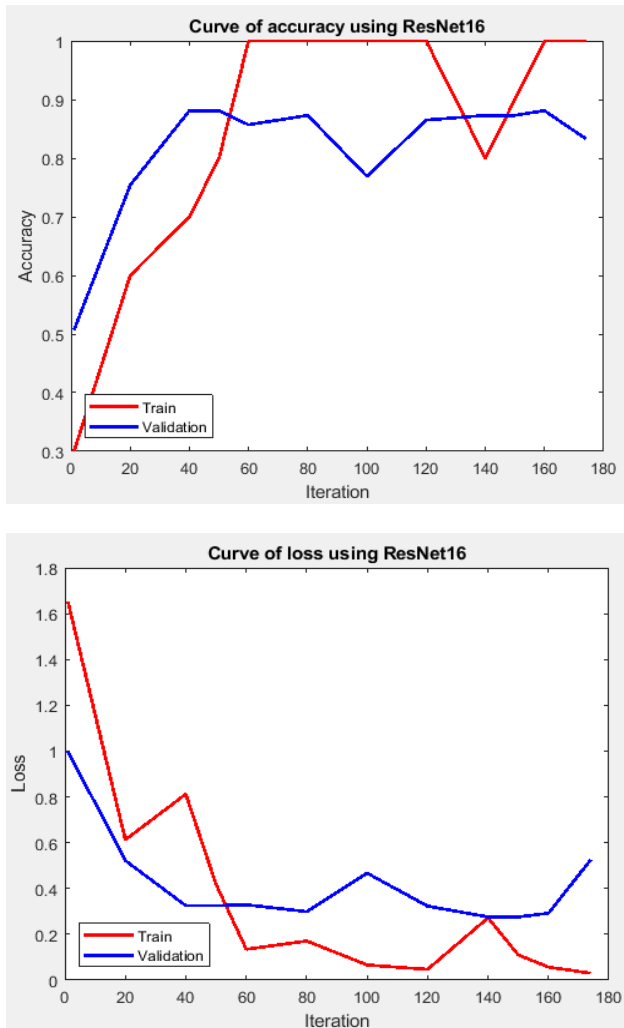

Figure 7. Obtained results using ResNet16

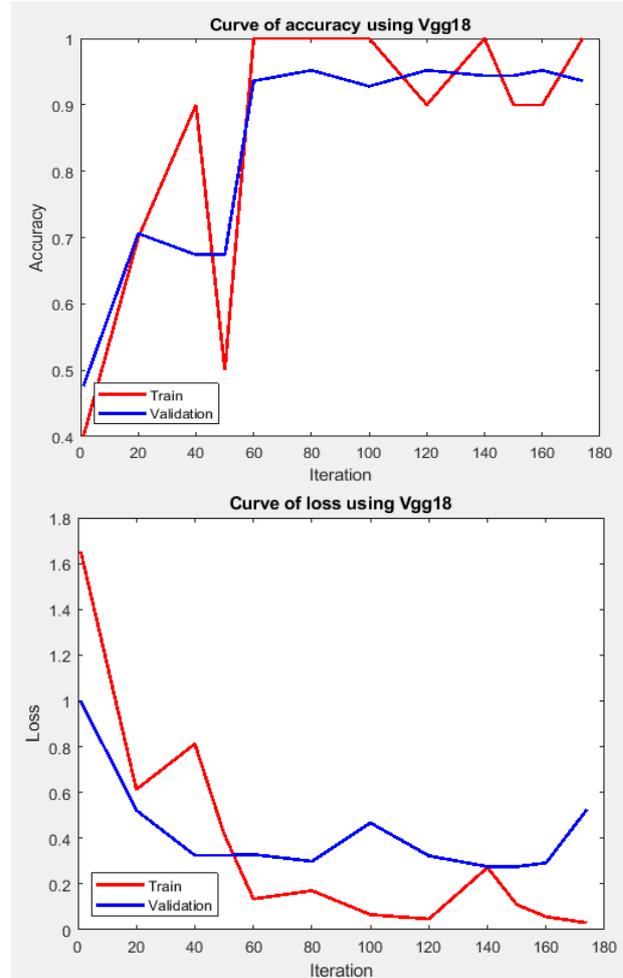

Figure 8. Obtained results using Vgg18

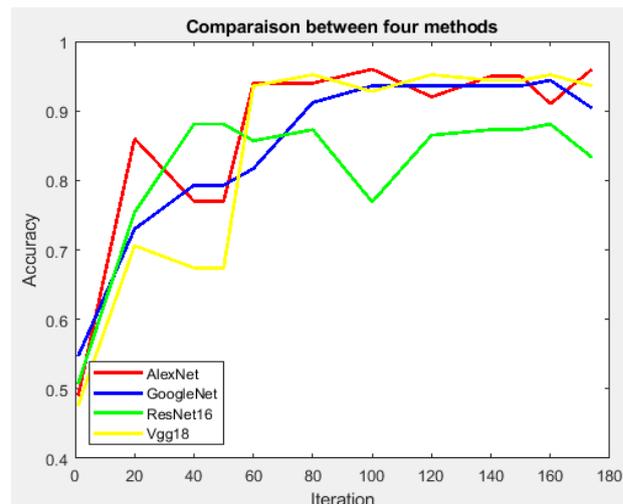

Figure 9. Comparison between four methods: AlexNet, GoogleNet, ResNet16, Vgg18

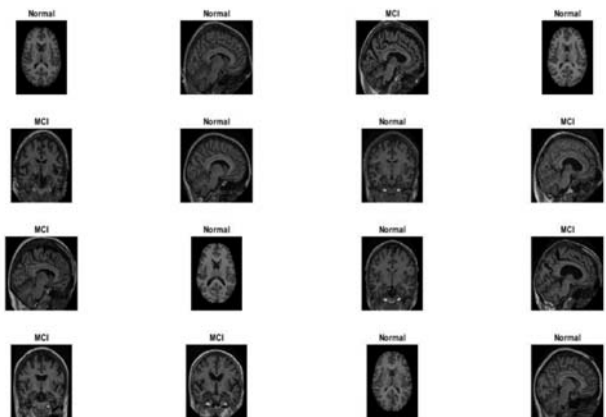

Figure 10. Results of some predictive subjects

## 5. Conclusion and perspectives

In this paper, we have introduced a new method for detecting AD at the MCI stage. The proposed method is based on a powerful classification architecture that implements the pre-trained network AlexNet to automatically extract the most prominent Magnetic Resonance Imaging (MRI) image's features. It can also serve as a starting point for improving accuracy in the diagnosis of other dementia such as Parkinson's disease. The important advantage of this new method is that early diagnosis of Alzheimer's disease can allow treatments and interventions before the loss of brain volume. As future work, we plan to train the algorithm in search of other models associated with the accumulation of beta-amyloid proteins and tau, two specific markers of Alzheimer's disease.

## Acknowledgments

This project was funded by Deanship of Scientific Research, Northern Border University for their financial support under grant no. SCI-2019-1-10-F-8183. The authors, therefore, acknowledge with thanks DSR technical and financial support.